# A link between solar events and congenital malformations: Is ionizing radiation enough to explain it?

Andrew C. Overholt[1], Adrian L. Melott[2], and Dimitra Atri[3]

[1] Department of Science and Mathematics, MidAmerica Nazarene University, 2030 East College Way, Olathe, Kansas 66062
[2] Department of Physics and Astronomy, University of Kansas, Lawrence, Kansas 66045
[3] Blue Marble Space Institute of Science, 1200 Westlake Ave N Suite 1006, Seattle, Washington 98109

ABSTRACT
  Cosmic rays are known to cause biological effects directly and through ionizing radiation produced by their secondaries. These effects have been detected in airline crews and other specific cases where members of the population are exposed to above average secondary fluxes. Recent work has found a correlation between solar particle events and congenital malformations. In this work we use the results of computational simulations to approximate the ionizing radiation from such events as well as longer term increases in cosmic ray flux. We find that the amounts of ionizing radiation produced by these events are insufficient to produce congenital malformations under the current paradigm regarding muon ionizing radiation. We believe that further work is needed to determine the correct ionizing radiation contribution of cosmogenic muons. We suggest that more extensive measurements of muon radiation effects may show a larger contribution to ionizing radiation dose than currently assumed.

1. INTRODUCTION

Beginning with the discovery of cosmic ray secondaries during the 1950s, researchers have been concerned with their effect on terrestrial biota. The work of these researchers has often focused on effects on air crews who spend long periods of time within the stratosphere where neutron secondaries are plentiful [*Reitz*, 1993; *O'Brien et al.*, 1996; *Sigurdson and Ron*, 2004; *Kojo et al.*, 2005; *Hammer et al.*, 2009; *Beck*, 2009; *among others*]. Biological effects are important at high altitude for cosmic rays of all energy ranges. However, biological effects at sea level have been found to be dominated by higher energy primaries, which produce more secondaries that may reach there. Recent work on this subject has suggested that lower energy events produced by the Sun may also produce biologic effects. The work of *Juckett* [2009] finds a correlation between birth defects, cancer, and the solar cycle. The work of *Belisheva et al.* [2012] as well as *Gromozova et al.* [2010] and *Gromozova et al.* [2012] (among others) focuses on solar events which produce increased neutron flux at ground level, called ground level events (GLEs). The work of *Belisheva et al.* [2012] finds similar results regarding congenital malformations (CMs). We consider the claims made in these works by simulating the total ionizing radiation flux at sea level due to solar events.

Ionizing radiation from astrophysical sources is dominated by cosmic ray secondaries of two forms: muons and neutrons. Neutrons in atmospheric matter are ejected through high energy collisions with cosmic rays and atmospheric nuclei. The energy needed for this spallation is low, with a neutron production threshold of a few MeV. Muons require much greater energies (> 1 GeV) for production in our atmosphere. These muons are produced through the decay of charged pions which are produced in high energy interactions in the upper atmosphere. These interactions are more rare, with fewer muons being produced than neutrons for a given cosmic ray primary. Despite their relative scarcity in comparison to neutrons, muons are a threat due to their high penetration. Muons are highly penetrating and often reach sea level or below, unlike cosmic ray neutrons [*Marinho et al.*, 2014]. Due to this penetration, muons produce the largest ionizing radiation dose of any secondary component from cosmic rays.

2. ENERGY DEPENDENCE OF GROUND LEVEL RADIATION

Ground level radiation varies greatly with primary spectrum. Neutrons are more plentiful at energies less than 1 GeV due to their low production energy threshold. Solar events are normally dominated by particles of this energy range. Muons have a much higher production threshold, greatly diminishing their contribution at the usual solar energy ranges. Figure 1 displays the ground level cosmic ray muon fluence to neutron fluence ratio at given primary energies. The data of this figure are calculated using the lookup tables of *Overholt et al.* [2013] and *Atri and Melott* [2011]. This ratio varies greatly with energy around the muon production threshold of ~1 GeV as shown.

Although electrons, protons, and other charged particles are produced in cosmic ray showers, these components do not contribute significantly to ground level radiation. These particles typically deposit their energy high in the atmosphere through ionization of atmospheric matter. The only charged particles which contribute significantly to ground level radiation are muons, due to their high penetration. Neutrons penetrate further in our atmosphere than charged particles due to their lack of energy loss due to ionization of atmospheric matter, increasing their abundance on the ground in comparison to charged particles. For this reason we focus on neutrons and muons.

3. DOSE CONTRIBUTION

The radiation dose from cosmic ray secondaries contains contributions from both muons and neutrons. Neutron radiation dose was found using the work of *Alberts et al.* [2001]. This dose was estimated by taking order of magnitude averages of ionizing radiation dose per neutron from *Alberts et al.* [2001] and multiplying by the number of neutrons found from the tables of *Overholt et al.* [2013]. Muon radiation dose was found by calculating the energy loss of high energy muons traveling through matter. Assuming a radiation-weighting factor of 1, this provides an equivalent dose. Research suggests that this assumption is adequate for deeper organs, but greatly underestimates dose at skin level [*Siiskonen,* 2008]. For these calculations, we use the energy loss of muons traveling through water as an approximation for biologic matter [*Klimushin et al.*, 2001]. Figure 2 displays the results of these calculations, showing the radiation dose of neutron and muon radiation at ground level due to different primary cosmic ray energies. Two irregularities exist within both Figures 1 and 2 corresponding to primary energies of ~$10^{9.3}$ *eV* and ~$10^{12.1}$ *eV*. These irregularities exist due to muon production efficiency at these energies and do not appear within the neutron data. The irregularity at about 3 GeV exists due to a resonance at the beginning of efficient muon production. The irregularity at about 2 TeV is not understood, but it has arisen in in experimental data from primaries at this energy [*Mauri and Sioli,* 2012] and is therefore unlikely to be a statistical fluctuation.

4. HISTORIC EVENTS

Using the neutron tables of *Overholt et al.* [2013] and the muon tables of *Atri and Melott* [2011], we now examine the integrated ionizing radiation dose from historic events. These tables provide the results of Monte Carlo simulations of cosmic ray interactions within our atmosphere. These tables were produced using CORSIKA for high energy atmospheric showers and Monte Carlo N-Particle (MCNP) Transport Code for low energy neutron thermalization. The details of these calculations can be found in the corresponding papers. The work of *Overholt et al.* [2013] provides neutron fluence per primary of given energy at different altitudes and energies within our atmosphere while the work of *Atri and Melott* [2011] provides muon fluence at various energies per primary of given energy at sea level. Convolving the cosmic ray spectra of solar events through these tables, we are provided with ground level fluences of muons and neutrons. We use the spectra of *Tylka and Dietrich* [2009] as our primary cosmic ray spectra. Figure 3 displays the spectra of the largest fluence events for which spectra exist. One of these events (October 1989) has also been linked to malformations in bacteria cultures in the work of

*Belisheva et al.* [2012]. The work of *Belisheva et al.* [2012] also finds a correlation between congenital malformations (CM) and the GLEs of this time period. After convolving the spectrum of *Tylka and Dietrich* [2009] through the lookup tables of *Overholt et al.* [2013], we find them to agree with the ground based neutron measurements of *Belisheva et al.* [2012] to within statistical uncertainty. The results of these calculations can be found in Table 1. As a method for checking these calculations, we have also calculated the radiation dose from the Feb. 23, 1956 GLE from increases detected by neutron monitors at the time [*Miroshnichenko et al.* 2013]. These increases were measured with neutron monitors active at the time of the GLE and measured as a total neutron flux increase above the background (measured in percent). Ground level neutron spectra can easily be found using total neutron flux from neutron monitor measurements. The work of *Overholt et al.* [2013] shows that the shape (but not the overall intensity) of low energy resolution ground-level neutron spectra remain independent of primary spectrum. As order of magnitude energy resolution neutron spectra are sufficient for radiation dose measurements, we will be using this for determination of the neutron radiation dose.

The work of *Juckett* [2009] does not focus on individual events, but rather the average fluence as evidenced by $\Delta^{14}C$ and neutron monitor measurements. As previously mentioned, total neutron flux increase is sufficient for this work. For the years covered by measurement of $\Delta^{14}C$ (1920-1954), we assume an increase in ground level neutron flux which is directly proportional to $\Delta^{14}C$. The assumption of a direct correlation between ground level neutron monitor measurements and $\Delta^{14}C$ is realistic, as the shape of the primary spectrum does not vary greatly during the time period examined. A drastic change in cosmic ray spectrum would be necessary to produce a ground level neutron flux which does not scale with $^{14}C$ production within our atmosphere. Specifically, it would require a very large change in low energy cosmic rays which could produce $^{14}C$ without any neutrons reaching ground level. Such a spectrum does not exist within the time period examined. Therefore, an increase in $\Delta^{14}C$ within the atmosphere should be accompanied by a directly proportional increase in ground level neutron flux. As a baseline association between cosmic ray spectrum and $^{14}C$ production we use the cosmogenic $^{14}C$ production of 2.2 $atoms\ cm^2\ s^{-1}$ and the spectrum during the 2010 solar minimum [*Kovaltsov et al.*, 2012]. For the years covered by neutron monitor measurements (1954-2000), we instead use directly measured neutron flux from monitors as given in *Juckett* [2009]. Recent results [*Smart et al.*, 2014] suggest that nitrate pulses detected in ice cores may indicate hard-spectrum events, but this is not yet developed in detail.

5. RESULTS

Our results show that the predominant source of ionizing radiation dose will be in the form of cosmic ray neutrons for solar events. The ratio of neutron radiation dose to muon radiation dose changes with the spectrum of the event, with softer spectrum events having a larger contribution from neutrons.

We find the average integrated dose from a single solar event varies greatly with event severity and spectrum, however is less than 1 μSv even in severe cases (~0.04 μSv for February

1956 event). This level of radiation is much less than even that of a small medical scan such as a foot x-ray (5 µSv), larger scans such as a CT chest scan (8 mSv), or a lethal dose (7 Sv) [*Wall and Hart,* 1997]. Research has shown that this level of radiation is insufficient for the production of birth defects [*ICRP* 2003]. This level of radiation is also much less than the annual average background radiation experienced by a member of the public in the United States (6.24 mSv) [*Ionizing radiation exposure of the population of the United States,* 2009].

For long-term increases in cosmogenic radiation referenced by *Juckett* [2009], we find the dose to be larger. The $\Delta^{14}C$ increase measured indicates a total cosmogenic radiation increase of at most ~10%. This only increases the average annual cosmogenic radiation dose at ground level by 0.03 mSv based upon the current average dose [*Ionizing radiation exposure of the population of the United States,* 2009]. Although this amount of radiation is larger than that produced in a solar event, it is still insufficient for a measurable increase in CM rate [*ICRP* 2003]. This exposure is also less likely to create CMs if the fluence is applied over a longer time period. This is due to repair mechanisms having a greater time to act in longer exposure periods [*Budworth et al.* 2012]. Neutron measurements referenced by *Juckett* [2009] give similar results to those described above.

A different situation exists with respect to the time period 773-776 AD, when a strong $\Delta^{14}C$ increase was found [*Miyake et al.* 2012]. This is most plausibly interpreted as an extreme solar particle event [*Melott and Thomas* 2012; *Usoskin et al.* 2013], or as a superposition of several SPEs of the October 1989 type [*Miroshnichenko and Nymmik*, 2014]. This event must have been a very hard-spectrum event in order to avoid unobserved strong atmospheric and biological effects [*Thomas et. al.*, 2013] Recently, high time-resolution data [*Ding et al.*, Method of high resolution on $^{14}C$ dating in coral in China and its implication on the $^{14}C$ spike event in AD 774 submitted to *Nuclear Instruments and Methods in Physics Research Proceedings* 2015] indicates that the increase was 45% over two weeks. We again use the approximation of linearity between $\Delta^{14}C$ and neutron flux. To obtain the muon dose associated with this event, we assume an event spectrally similar to the largest recorded SPE (Feb. 23 1956), though with much greater magnitude. These results (displayed in Table 1) produce a scenario of greatly increased radiation dose, at an integrated dose of 16 µSv. However, even this level of radiation is below the amount of ionizing radiation dose of medical diagnostic procedures which research has shown does not increase risk of congenital malformation [*ICRP* 2003].

6. DISCUSSION

The works of *Belisheva et al.* [2012] and *Juckett* [2009] find correlations between cosmic ray flux and CMs. These correlations suggest that a very small increase in ionizing radiation dose could increase the CM rate of a population.

Here we have calculated the cosmogenic radiation dose due to the two largest contributing particle species at ground level. These calculations find that ionizing radiation alone is insufficient for the production of CMs for solar events. Although it is possible that the slight increase in ionizing radiation experienced during a solar event could increase the probability of

CMs, we find that the detected increase in CM rate is too large to be produced by the increase in ionizing radiation dose calculated. This is primarily due to repair mechanisms which allow cells to repair damage due to small amounts of radiation [*Budworth et al.* 2012].

One possible explanation resides in the DNA damage due to cosmogenic muons. Although the effects of neutron radiation on the human body are well known, the same is not true for muons. Muon radiation is assumed to be equivalent to other ionizing radiation sources (a radiation-weighting factor of 1) as experiments finding the DNA damage due to muons have not been attempted. The work of *Siiskonen* [2008] finds this assumption to be a substantial underestimate in the case of surface tissue, and claims this assumption could be true for deeper organ tissue. The bystander effect suggests that a radiation-weighting factor of 1 may not be correct for deeper organ tissue. Experiments have displayed the bystander effect by irradiating 10% of a cell group. These cell groups show identical damage to groups which were 100% irradiated [*Sawant et al.,* 2001]. As muons penetrate deeply into the body, even a small flux of muons could irradiate a large group of cells deep within the body. This could explain why the bystander effect would be more prevalent in muon radiation as opposed to other forms of less penetrating radiation.

Another likely possibility is that the effects of muons and neutrons together are greater than either form of radiation by itself. It has been shown that two genotoxic events of similar biological and mechanistic behavior produce effects which are additive [*UNSCEAR* 2000]. Although the mechanistic behavior of muon and neutron radiation are similar (and would thus be additive), they distribute radiation very differently within tissues. Muons are sparsely ionizing and would thus spread damage over a very large region. In contrast, the damage created by high-LET radiation such as neutrons is localized to the path of the particle. These two effects will combine in the case of cosmic radiation and thus could account for more damage than ionization alone suggests.

To verify this theory, experiments must be performed which irradiate tissues with muons alone and in combination with neutrons and measure the resulting DNA damage. The correlations documented by *Juckett* [2009] and *Belisheva et al.* [2012] pose a mystery in the light of this research. To explain these correlations, we must re-examine our assumptions about the damage caused by cosmic ray secondaries. Although neutron radiation damage has been well studied in the past, muon radiation damage remains a large question which may hold the answer to this mystery.

Acknowledgements
We acknowledge the useful comments from Andy Karam, two anonymous referees, and support from NASA grant no. NNX14AK22G. All data used in this research are made freely available at:
http://kusmos.phsx.ku.edu/~melott/crtables.htm

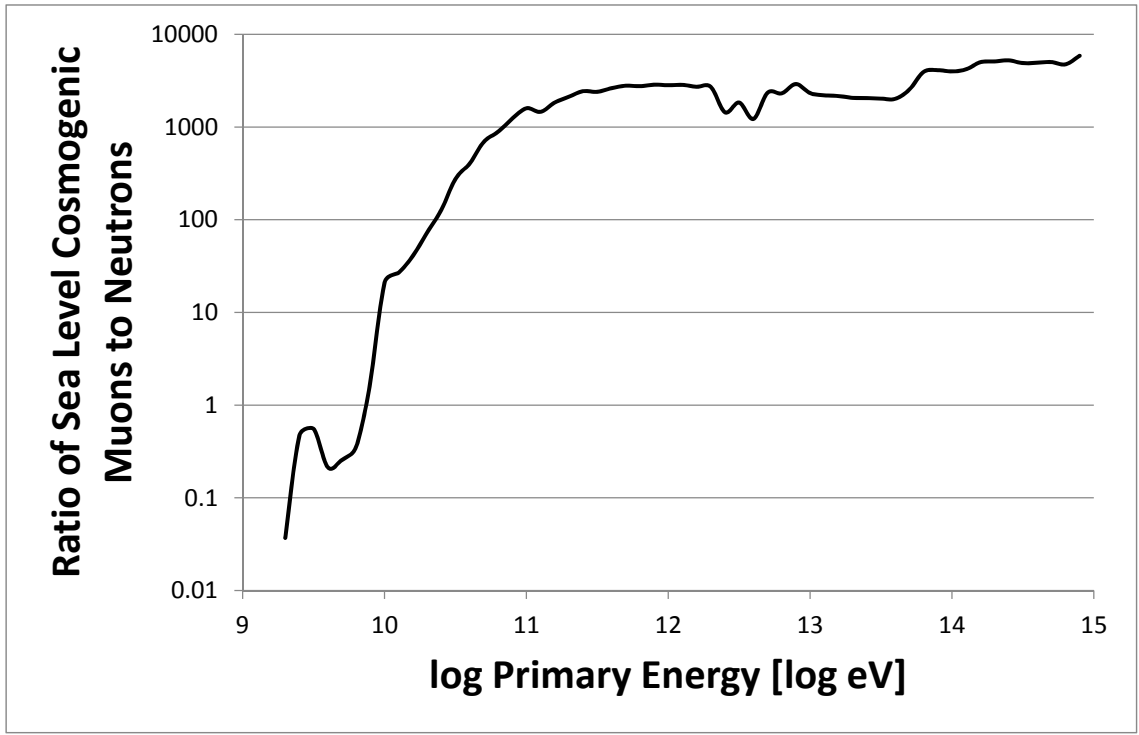

**Figure 1.** Ratio of the number flux of cosmogenic muons to neutrons which reach sea level from primaries with energies from 1 GeV to 1 PeV. Neutrons have a low threshold for production energy, making them much more plentiful at low energies. The threshold for muon production is higher, thus they become more numerous at primary energies above 10 GeV. Neutron and muon flux at sea level is insubstantial from primaries below 1 GeV.

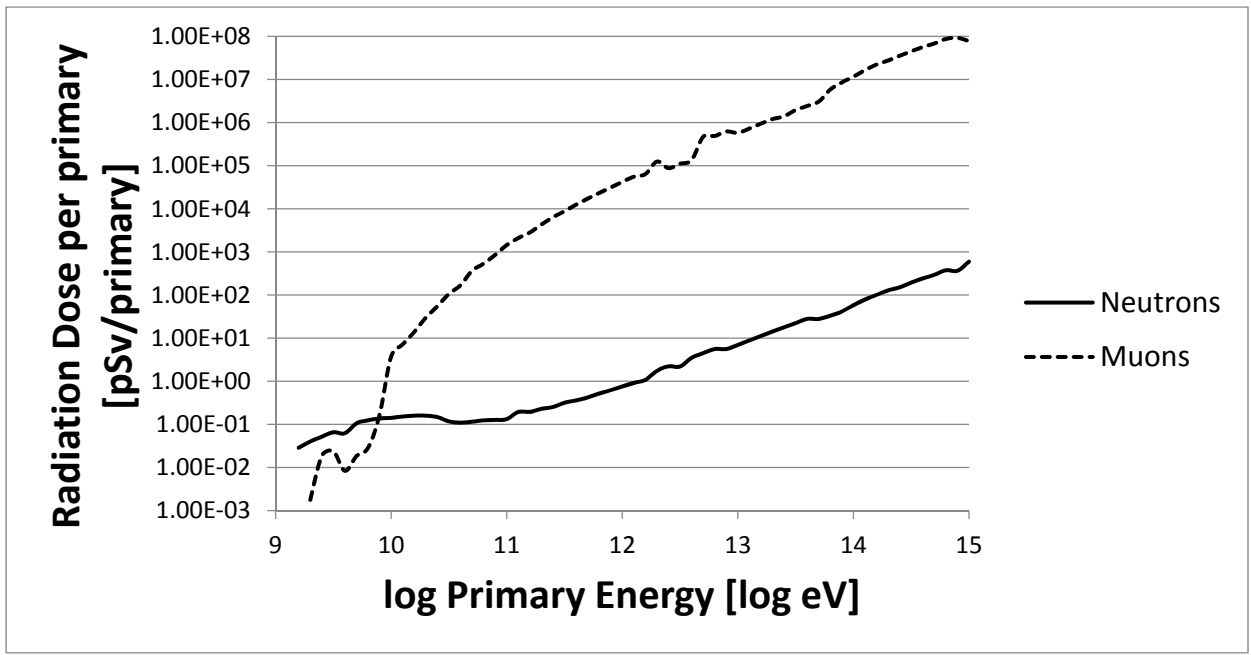

**Figure 2.** Dose at sea level due to different cosmic ray secondaries for primary energies from 1 GeV to 1 PeV. Though the biological effectiveness of individual neutrons is greater than that of muons, the large number of muons produced by high energy cosmic rays make them the dominant source of radiation above 10 GeV.

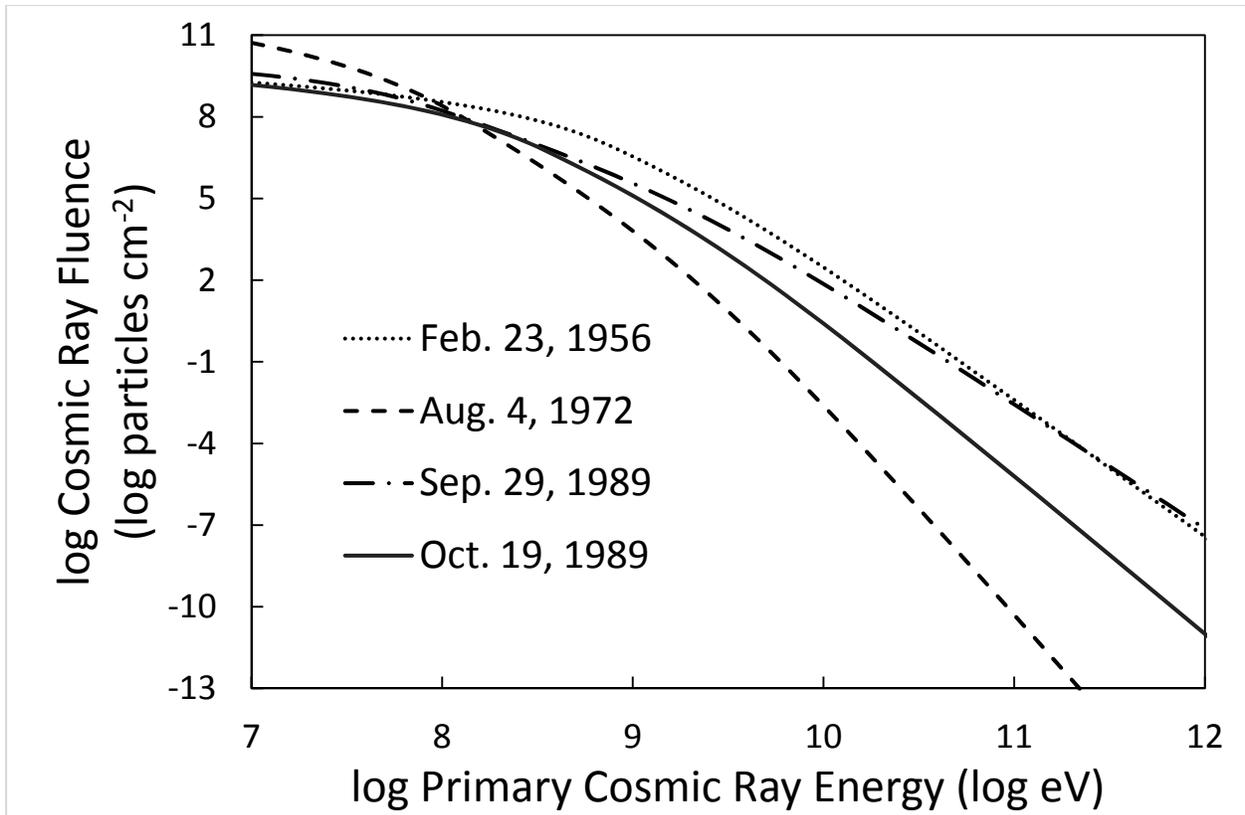

**Figure 3.** Cosmic ray fluence for the largest events on record: Feb. 23, 1956; Aug. 4, 1972; Sep. 29, 1989; and Oct. 19, 1989 calculated in the present work from the spectra of *Tylka and Dietrich* [2009].

**Table 1.** Ionizing radiation dose at sea level from historic SPEs.

| Event Date | Neutron Dose (µSv) | Muon Dose (µSv) | Total Dose (µSv) |
|---|---|---|---|
| February 23, 1956 | 0.043 | 0.007 | 0.050 |
| August 4, 1972 | 0.000022 | 0.0000010 | 0.000023 |
| September 29, 1989 | 0.0057 | 0.0016 | 0.0073 |
| October 19, 1989 | 0.0061 | 0.0020 | 0.0081 |
| October 22, 1989 | 0.000074 | 0.0000043 | 0.000078 |
| October 24, 1989 | 0.0017 | 0.00019 | 0.0019 |
| 773-776 AD | 14 | 2.2 | 16 |